\documentclass[reprint,amsmath,amssymb,aip,floatfix,superscriptaddress]{revtex4-1}
\usepackage{graphicx}
\usepackage{bm}
\usepackage{amsmath}
\usepackage[colorlinks=true,citecolor=blue,linkcolor=blue,urlcolor=blue]{hyperref}
\usepackage{epstopdf}
\begin{document}

\title{Perfect valley polarization in MoS$_2$}

\author{A. Heshmati-Moulai}
\affiliation{Department of Physics, Iran University of Science and Technology, Narmak, Tehran 16844, Iran}

\author{H. Simchi}
\email{simchi@alumni.iust.ac.ir}
\affiliation{Department of Physics, Iran University of Science and Technology, Narmak, Tehran 16844, Iran}
\affiliation{Semicondutor Technology Center, Tehran 19575-199, Iran}

\author{M. Esmaeilzadeh}
\email{mahdi@iust.ac.ir}

\affiliation{Department of Physics, Iran University of Science and Technology, Narmak, Tehran 16844, Iran}

\date{\today}

\begin{abstract}
We study perfect valley polarization in a molybdenum disulfide (MoS$_2$) nanoribbon monolayer using two bands Hamiltonian model and non-equilibrium Green's function method. The device consists of a one-dimensional quantum wire of MoS$_2$ monolayer sandwiched between two zigzag MoS$_2$ nanoribbons such that the sites A and B of the honeycomb lattice are constructed by the molecular orbital of Mo atoms, only. Spin-valley coupling is seen in energy dispersion curve due to the inversion asymmetry and time-reversal symmetry. Although, the time reversal symmetry is broken by applying an external magnetic field, the valley polarization is very small. A valley polarization equal to 46\% can be achieved using an exchange field of 0.13 eV. It is shown that a particular spin-valley combination with perfect valley polarization can be selected based on a given set of exchange field and gate voltage as input parameters. Therefore, the valley polarization can be detected by detecting the spin degree of freedom.
\end{abstract}

\pacs{73.63.-b, 75.70.Tj, 78.67.-n, 85.35.-p}

\keywords{MoS$_2$, Valley polarization, Time-reversal symmetry, Inversion symmetry, Non-equilibrium Green's function}

\maketitle

\section{Introduction}
Semiconductors changed the world beyond anything that could have been imagined before. Rectifiers, field effect transistors, tunneling diodes, integrated chips, and lasers are among such semiconductor products. Although the history of the semiconductor devices is long and complicated, the electron charge injection plays the essential and main rule in the field. By miniaturization of the devices, not only the power consumption and manufacturing expenses increase but also the repeatability of manufacturing process would be very difficult \cite{Haselman2010}. In consequence, spin degree of freedom (DOF) has been considered as the main parameter in device design. In comparison with the charge-based devices, the higher data processing speed and its non-volatility, lower power consumption and higher integration densities are the main competencies of spin-based devices \cite{awschalom2002spintronics}. Of course, the working temperature, the rate of spin injection and perfect spin polarization are the main bottlenecks of the spin-based components \cite{awschalom2002spintronics}. Therefore, scientists and technologists have been motivated to use the valley DOF for device design.

In two-dimensional materials, the valence and conduction bands become close to each other at $K$ and $K'$ points, the two inequivalent corners of the hexagonal Brillion zone.  The low-energy electron or hole near these points can be described by the Dirac equation. In consequence, the Dirac-cone-like band structure forms in the vicinities of the points called $K$ and $K'$ valleys (Dirac points). In other words, near the Dirac points an electron or hole possesses two more degrees of freedom called pseudospin or valley DOF \cite{torres2014introduction}. It has been shown that the wave packet in a magnetic Bloch band rotates and creates an orbital magnetic moment. The magnetic moment causes the valley orbital interaction (VOI) \cite{Chang1996}. Recher \textit{et al.}, have shown that, the combined effect of ring confinement and applied magnetic flux offers a controllable way to lift the valley degeneracy in graphene rings \cite{Recher2007Aharonov}. The valley dependent Berry phase effect which can produce a valley contrasting Hall transport has been studied \cite{Xiao2007}. Recher \textit{et al.}, have considered single-layer and bilayer graphene quantum dots and have shown that, the valley degeneracy is efficiently and controllably broken by applying a perpendicular magnetic field to the graphene plane \cite{Recher2009Bound}. It has been shown that, a finite length line defect superlattice in graphene can be utilized to realize valley-filtering function \cite{Xiao-Ling2012}. Wu \textit{et al.}, have studied the valley pair qubit in double quantum dots of gapped graphene \cite{Wu2012}. 

Generally, there are two prerequisites for the generation of the valley-polarized current: (i) to break the valley symmetry and (ii) to break the time reversal symmetry \cite{ Fujita2010,Yesilyurt 2016}. It has been shown that, the valley and time reversal symmetries are broken by applying a strain and magnetic field to graphene, respectively \cite{ Fujita2010, Yesilyurt 2016}. In addition, the use of potential barriers is one of the common ways to generate a filter function. When electrons impinge to the barrier the electrons whose incident angle is larger than the critical angle will be reflected and in consequence the conductance of the system decreases \cite{ Fujita2010, Yesilyurt 2016}. It has been shown that, the electrons of the undesired valley could be totally reflected by transverse shifting of the propagation direction, i.e., by using a ferromagnetic strip with magnetization up (down). After passing through the first barrier, the transmitted electrons move along a strained monolayer graphene sheet and impinge to the second barrier which is created by a ferromagnetic strip with magnetization down (up). It has been shown that, both the valley polarization and conductance are high in this structure \cite{Yesilyurt 2016}.

Silicene is a good candidate for spin-valleytronics due to its interesting electronic properties such as the controllability of Dirac mass by electric field, and the spin-valley-dependency in its band structure \cite{ Soodchomshom 2014}. Zigzag silicene nanoribbon is metal\cite{shakouri2015tunable}. When a finite length of the ribbon is sandwiched between two different magnetic insulators, silicene-based normal metal/sublattice-dependent ferromagnetic/normal metal (NM/SFM/NM) junction is formed. It has been shown that, a perfect spin-valley polarization can be seen when an electric field is applied perpendicular to the plane of Silicene\cite{ Soodchomshom 2014}. Yesilyurt et al., have used a uniform uniaxial strain on the Silicene lattice instead of an electric field \cite{Yesilyurt 2015}. There are two regions in their structure. The first region is sandwiched between two different magnetic insulators. The region is utilized to create an angular separation in the transmission of different spin and valley currents. The second region consists of two asymmetric FM stripes and electric potential barrier induced by top and bottom gates. They have demonstrated controllable and highly-efficient filtering (exceeding 90\%) for all four spin-valley combinations based on realistic parameter values \cite{Yesilyurt 2015}. 

Finally, Shan et al., have studied the effect of impurities and disorders on the extrinsic spin Hall conductivity (SHC) in spin-valley coupled monolayers of transition metal dichalcogenides \cite{Shan 2013}. They have applied the standard diagrammatic approach, in which the scattering due to impurities and disorders is treated as the perturbation to the eigenstates
of Hamiltonian\cite{Shan 2013}.

The monolayer MoS$_2$ has a honeycomb lattice with potential applications in two-dimensional nano-devices \cite{Kuc2011,Li2012}. Since MoS$_2$ is a direct band gap semiconductor, it is suitable for optical manipulations and opens access to many optoelectronic applications \cite{Kuc2011,Li2012,Cao2012}. Xiao \textit{et al.}, have shown that the spin and valley are coupled in monolayers of MoS$_2$ due to the inversion asymmetry and the presence of spin-orbit interaction \cite{xiao2012coupled}. Also, They have studied the optical interband transition by using a $\vec{k}.\vec{p}$ Hamiltonian model. Cappelluti \textit{et al.}, have used the Slater-Koster method and described the energy dispersion curve of the monolayer of MoS$_2$ by using an eleven-band Hamiltonian model \cite{Cappelluti2013}. It has been shown that in zigzag nano-ribbon of  MoS$_2$, there is a mismatch between valley Zeeman coupling in valence and conduction bands due to the effective mass asymmetry effect which is proportional to the square of applied magnetic field intensity \cite {rostami2014intrinsic}. Also, the spin polarization has been studied when there are random crystal defects and impurities in the nanoribbon in the presence of a high intensity magnetic field  \cite {rostami2014intrinsic}. Rostami \textit{et al.}, have studied the variations of electronic properties of the monolayer of MoS$_2$ under different strain conditions \cite{rostami2015theory}. In continuance of previous works \cite{Cappelluti2013,rostami2015theory,rostami2014intrinsic}, Rostami \textit{et al.}, have shown that, the edge effects exist in zigzag nanoribbon of the monolayer of MoS$_2$ due to the edge states and the monolayer of MoS$_2$ is a valley Hall insulator \cite{rostami2015edge}. Also, the effect of spatially modulated magnetic field on the electronic properties of the monolayer of MoS$_2$ has been studied by Li \textit{et al.} \cite{Li2016}. Therefore, for describing the electronic structure and properties of MoS$_2$, many first principles and tight binding studies have been done up to now  \cite{Cappelluti2013,rostami2015theory,rostami2014intrinsic,Rostami2015b,rostami2015edge,Wolfram2014,Vafek2014,Li2016}.

In continuance of the mentioned previous works, we consider a monolayer MoS$_2$ zigzag nanoribbon in 1H structural phase and study the valley-polarization and valley-selective transport using two bands Hamiltonian model and non-equilibrium Green's function method. We assume, the sites A and B of the honeycomb lattice are constructed by the molecular orbital of Mo atoms, only  \cite{Tahir2016Quantum,Klinovaja2013Spintronics}. We show that the time reversal symmetry is broken by applying an external magnetic field. Under this condition, the valley polarization is very small. But, by applying an exchanged field of 0.13  eV, a valley polarization equal to 46\% is achieved.We show; a particular spin-valley combination with perfect valley polarization can be selected based on a given set of exchange field and gate voltage as input parameters.  The structure of the article is as follows: in section \ref{sec:basicformalism}, device structure, the $\vec{k}.\vec{p}$ Hamiltonian model and non-equilibrium Green's function method are illustrated. The numerical results and discussion are provided in section \ref{sec:numerical} and a summary is given in section \ref{sec:summary}.

\section{Basic formalism}\label{sec:basicformalism}

\subsection{Device structure}
We consider a monolayer zigzag nanoribbon whose finite length (called channel) is sandwiched between two magnetic insulators (e.g., EuO). The exchange energies induced by the top and the bottom magnetic insulators into A- and B-sublattices are $h_{1}$ and $h_{2}$, respectively and the chemical potential  is induced by the top and the bottom gates with the same potential $V_{G}$  \cite{ Soodchomshom 2014}. The sites A and B of the honeycomb lattice are constructed by the molecular orbital of Mo atoms, only \cite{Tahir2016Quantum,Klinovaja2013Spintronics}. In addition, the difference between Fermi energy ($E_{F}$) of leads and channel is equal to $(h_{1}+h_{2})/2+V_{G}$ (see Fig.1). 

\subsection{$\vec{k}.\vec{p}$ Hamiltonian model}
The low-energy electronic states are mainly dominated by $(4d_{3z^2-r^2},4d_{xy},4d_{x^2-y^2})$ orbitals of Mo atoms indexed by the magnetic quantum number $m_l=0,-2,2$, respectively\cite{Li2016}. By considering $| \phi_c \rangle=| d_{3z^2-r^2}\rangle$ and $| \phi^{\tau}_{v} \rangle=| d_{x^2-y^2}+i \tau d_{xy} \rangle$ as the basis wave vectors, where the subscript $c(v)$ indicates conduction (valence) band, and $\tau=\pm1$ is the valley index, the two band $\vec{k}.\vec{p}$ Hamiltonian has the following form \cite{xiao2012coupled,kormanyos2015,Tahir2016Quantum,Klinovaja2013Spintronics}

\begin{figure*}
\includegraphics[width=0.4\linewidth]{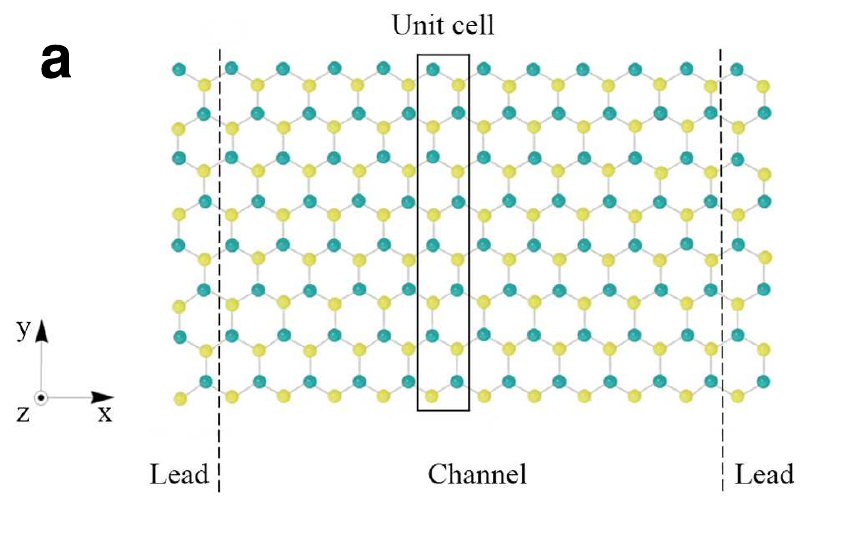}
\hspace{20pt}
\includegraphics[width=0.44\linewidth]{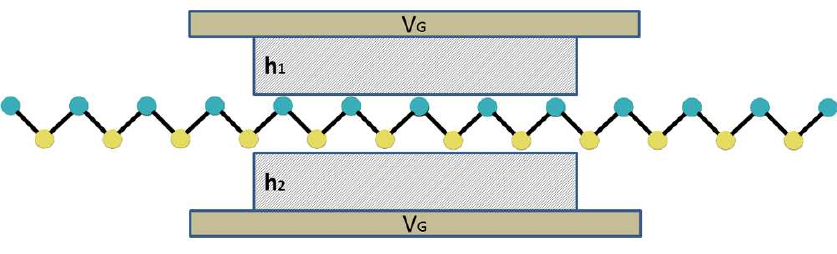}
\caption{(Color online) (a) A zigzag-like nanoribbon. Each unit cell includes 8 sites A (representing ${d_{3{z^2} - {r^2}}}$) and 8 sites B (representing ${d_{{x^2} - {y^2}}} + i\tau {d_{xy}}$ ). Site A and site B are shown in green and yellow color, respectively. (b) Device structure including zigzag nanoribbon, magnetic insulators ($h_{1}$ and $h_{2}$) and top and bottom gates with potential $V_{G}$.}
\label{fig:Fig1}
\end{figure*}

\begin{widetext}
\begin{equation}\label{eq:2bandH}
H =  
\begin{pmatrix}
\Delta + s{M_z} - \tau {M_\tau } - sM + {eV_G}    &    at\left( {\tau {k_x} - i{k_y}} \right)    \\
at \left( \tau {k_x} - i{k_y} \right) &     - {\rm{\Delta }} + 2\lambda \tau s + s{M_z} - \tau {M_\tau } - sM + {eV_G}
\end{pmatrix}
\end{equation}
\end{widetext}

where, $\lambda  = 0.0375$ eV is the spin-orbit coupling constant, ${V_G}$ is the gate voltage, $s =  \pm 1$ is the spin index, ${\rm{\Delta }} = 0.83$ eV  is the energy gap (mass term), $a=3.2$~\AA~ is the lattice constant, and $t=1.27$ eV is the hopping integral. The terms containing ${M_z} = 2.21{\mu _B}B/2$ and ${M_\tau } = 3.57{\mu _B}B/2$ correspond the regular Zeeman and valley Zeeman exchange fields, of which the latter breaks the valley symmetry of the levels \cite{Tahir2016Quantum} with ${\mu _B} = 5.788 \times {10^{ - 5}}$ eV/T. Finally, $M$ is the strength of the external exchange field. By considering a honeycomb lattice composed of two kinds of atoms, A (${d_{3{z^2} - {r^2}}}$) and B (representing ${d_{{x^2} - {y^2}}} + i\tau {d_{xy}}$ ), an effective tight binding Hamiltonian is found as follows\cite{Klinovaja2013Spintronics}

\begin{widetext}
\begin{equation}\label{eq:effectiveH}
{\bar H_0} = 
\begin{pmatrix}
{{\rm{\Delta }} + s{M_z} - \tau {M_\tau } - sM + {eV_G}}&t\\
t&{ - {\rm{\Delta }} + 2\lambda \tau s + s{M_z} - \tau {M_\tau } - sM + {eV_G}}
\end{pmatrix}
\end{equation}
\end{widetext}

The Hamiltonian $H$ and $\bar H_0 $ are valid for {color{red}momentum} close to $ \pm K$ and result in the same low-energy spectrum \cite{Tahir2016Quantum,Klinovaja2013Spintronics}. The conduction band minimum (CBM) is the reference point of energy in both Eqs.~\ref{eq:2bandH} and \ref{eq:effectiveH} and $\hbar\vec k \equiv ({\hbar k_x},{\hbar k_y})$ is the crystal momentum.

The valley-dependent conductance and valley polarization are studied using tight binding non-equilibrium Green's function method (TB-NEGF). The used TB-NEGF method has been explained in our previous work in detail \cite{shakouri2015tunable}. The valley polarization is defined as

\begin{equation}\label{eq:valleyPol}
{P_v} = \frac{{I_{\tau=1}^{s =+1} + I_{\tau=1}^{s =-1} - I_{\tau=-1}^{s=+1} - I_{\tau=-1}^{s =-1}}}{{I_{\tau=1}^{s=+1} + I_{\tau=1}^{s=-1} + I_{\tau=-1}^{s=+1} + I_{\tau=-1}^{s =-1}}}
\end{equation}

where $I_\tau ^s$ is the current of the electrons with spin $s$ and valley $\tau$. The current is calculated using the below formula 

\begin{equation}\label{eq:currentformula}
{I}= \frac{2e}{\hbar} \int_{-\infty}^{+\infty}{dE}{G(E)}{({f_{1}{(E)}}-{f_{2}{(E)}})}
\end{equation}
Or (due to the properties of Fermi distribution function)
\begin{equation}\label{eq:currentformula2}
  I= \frac{2e}{\hbar} \int_{\mu_{1}}^{\mu_{2}}{dE}{G(E)}{({f_{1}{(E)}}-{f_{2}{(E)}})}
\end{equation}
where, {\color{red} $G(E)$} is transmission probability and $\mu_{i}$ and $f_{i}$ are Fermi energy and Fermi distribution function, respectively \cite{SDatta2005}.The Fermi distribution function depends on the temperature $T$ as $f(E)=1/(e^{(E-E_{f})/KT}-1)$ where $K$ is Boltzmann constant and $E_{f}$ is Fermi energy. We will consider $T=300$ Kelvin in all next calculations.

\begin{figure*}
\includegraphics{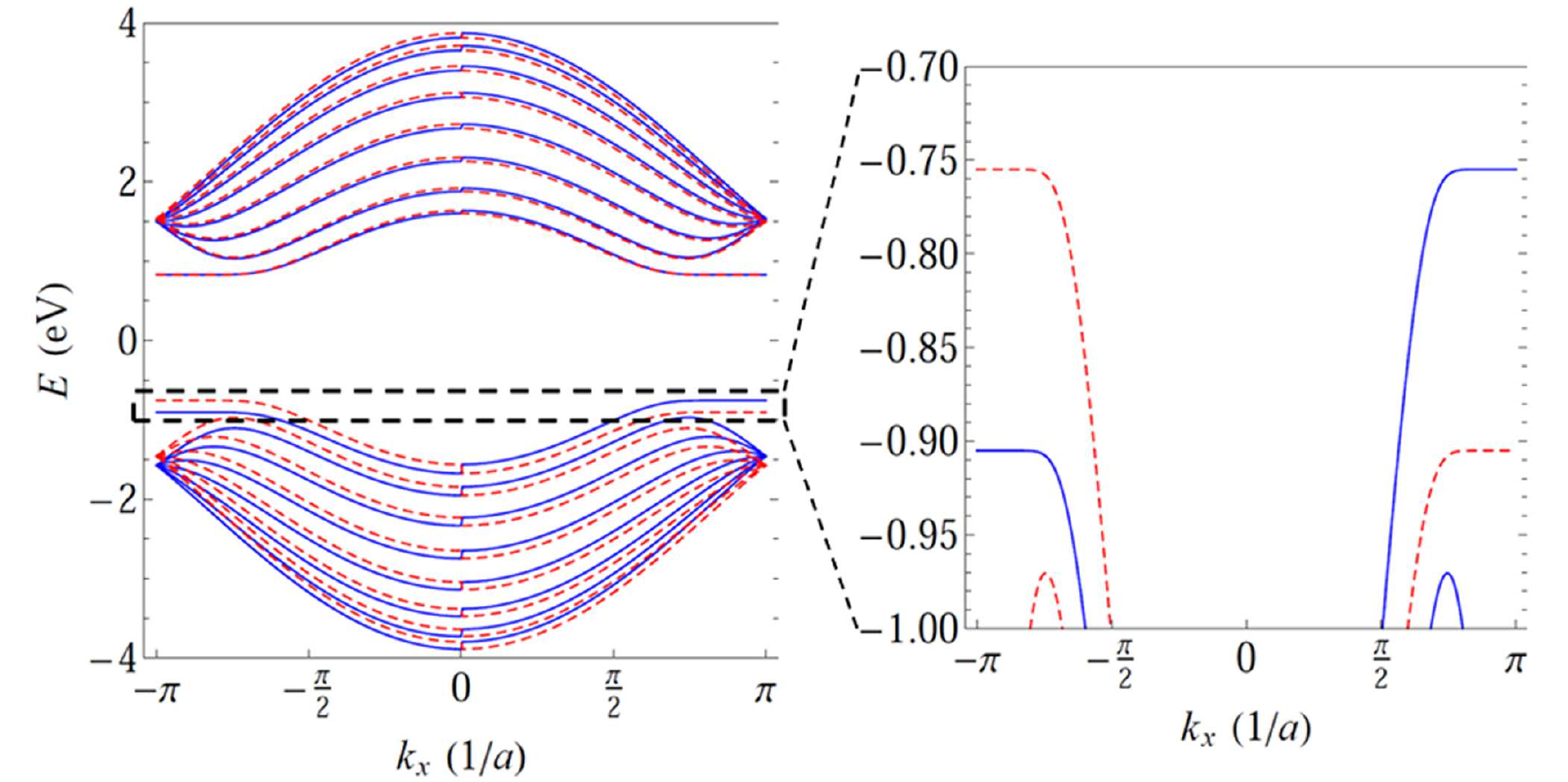}
\caption{(Color online) Energy dispersion plot of zigzag MoS$_2$ nanoribbon for energy range $-4\le E \le 4$ eV. The inset shows the range  $-1\le E \le -0.75$ eV. Spin-up bands are in blue and spin-down bands are depicted in red. Conduction band is degenerate but valence band is not. The energy gap is equal to $2\Delta=1.66$ eV based on the $\vec k.\vec p$ model. The $K$ ($K'$) point corresponds to $\tau=+1(-1)$. Strong valley-spin coupling is seen in the valence band due to the inversion asymmetry.}
\label{fig:Fig2}
\end{figure*}

\section{Numerical results and discussion}\label{sec:numerical}
The schematic of a zigzag nanoribbon is shown in Fig.~\ref{fig:Fig1}. As the figure shows, the width and length of the ribbon are along the $y$ and $x$-directions, respectively and the $z$-direction is perpendicular to the $xy$-plane. Below, we consider three different cases, which are the Hamiltonian without the effects of the magnetic and the exchange fields, with the magnetic field and without the exchange field and finally, including the effect of exchange field and without the magnetic field.

\subsection{Without the magnetic and the exchange fields}
\begin{figure}
\includegraphics[width=.8\linewidth]{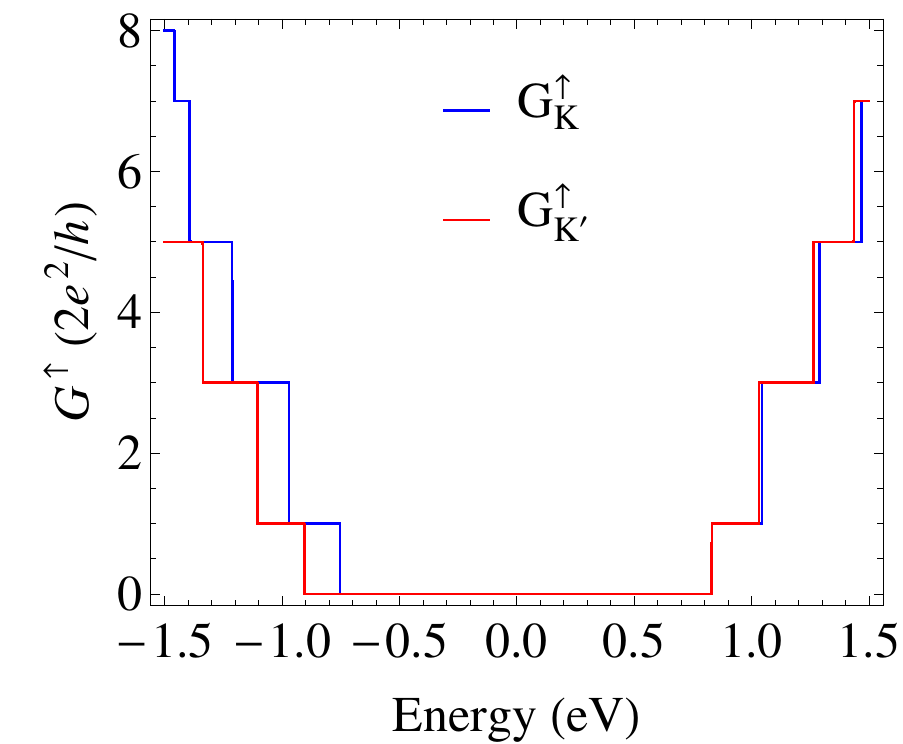}
\includegraphics[width=.8\linewidth]{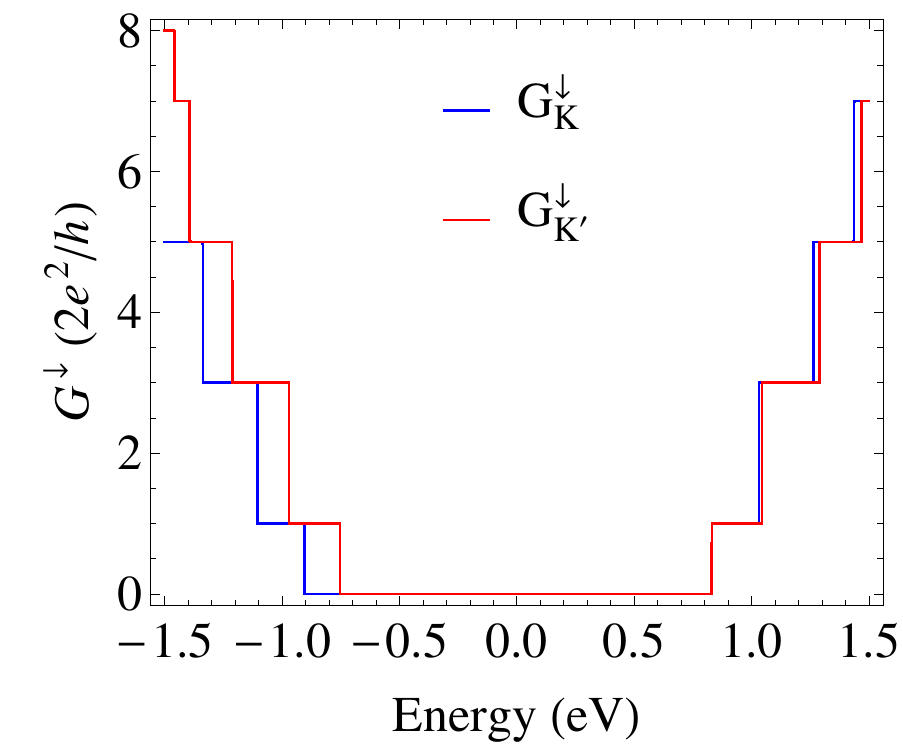}
\caption{(Color online) Conductance versus electron energy (a) spin-up ($\uparrow$) and (b) spin-down ($\downarrow$). The $K$ ($K'$) point corresponds to $\tau=+1(-1)$.}
\label{fig:Fig3}
\end{figure}

The massive Dirac Hamiltonian in two-dimensional materials (Eqs.~\ref{eq:2bandH} and \ref{eq:effectiveH}) implies the existence of energy band gap in $E(k)$ \cite{xiao2012coupled}. Fig.~\ref{fig:Fig2} shows the energy dispersion plot ($E(k)$) of the nanoribbon which includes an energy band gap equal to $2{\rm{\Delta }} = 1.66$ eV. The existence of $2\lambda \tau s$ term in Eqs.~\ref{eq:2bandH} and \ref{eq:effectiveH} reflects the strong valley-spin coupling in the valence band \cite{xiao2012coupled}. As Fig.~\ref{fig:Fig2} shows, the plot of $E(k)$ in valence band for $s =+1$ and $\tau=+1$ is similar to the case $s =-1$ and $\tau=-1$ due to the time reversal symmetry (TRS) in MoS$_2$. Also, the case $s =+1(-1)$ and $\tau=-1(+1)$ is similar to the case $s =-1(+1)$ and $\tau=+1(-1)$ due to the inversion symmetry in MoS$_2$. Therefore, a strong valley-spin coupling is seen in the valence band. The breaking of the valley degeneracy is evident in the inset of Fig. \ref{fig:Fig2} for $-0.9\le E \le -0.75$ eV. In the next calculations, we focus on this range of energy for studying the valley polarization in the device. The plot of the quantum conductance versus the energy includes a transmission gap, conduction and valence bands and significant valley-dependent conductance at the edge of the valence band due to the strong valley-spin coupling term in Hamiltonian (Fig.~\ref{fig:Fig3}). As Fig.~\ref{fig:Fig3}(a) and {Fig.~\ref{fig:Fig3}(b)} show, the conductance of spin-up (spin-down) electrons at the edge of the valence band at $K$($K'$)-point is equal to ${1G_0}$ approximately while at $K'$($K$)-point it is equal to zero. Therefore, the valley polarization is equal to zero at the edge of valence band due to the time reversal symmetry although for each type of spin it is equal to one due to the inversion asymmetry.

\subsection{With the magnetic field and  without the exchange field}
\begin{figure}
\includegraphics[width=.8\linewidth]{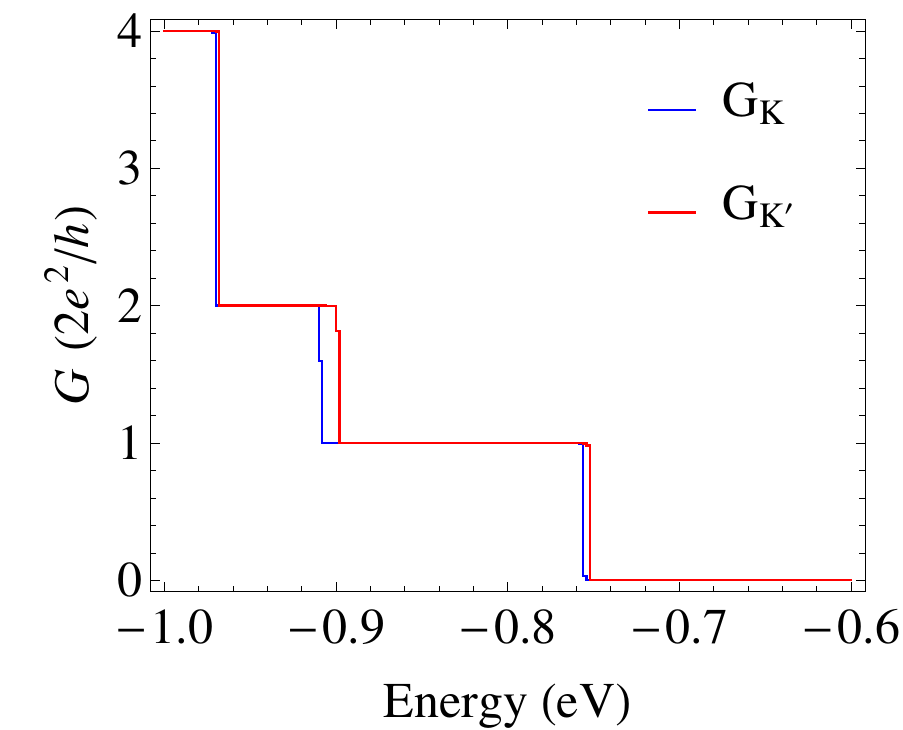}
\caption{(color online) Conductance versus electron energy in the energy interval $-1\le E\le -0.6$ eV, and $B_0=30$~T.}
\label{fig:Fig4}
\end{figure}

For inducing a valley polarization, we should break the time reversal symmetry. For doing it, we can apply a magnetic field ($\vec B = {B_0}\hat z$) perpendicular to the Mo-plane. Fig.~\ref{fig:Fig4} shows the total conductance ($G^\uparrow$+$G^\downarrow$) versus the electron energy for $B_0=30$ T and $-1\le E\le -0.6$ eV. As the figure shows, valley splitting occurs due to the breaking of the time-reversal symmetry. Of course, the pure valley conductance (defined by $G_K - G_{K'}$) is not significant despite the high value of the magnetic field. Therefore, we should use another technique for finding a perfect valley polarization.

\subsection{With the exchange field and without the magnetic field}
\begin{figure}
\includegraphics[width=.8\linewidth]{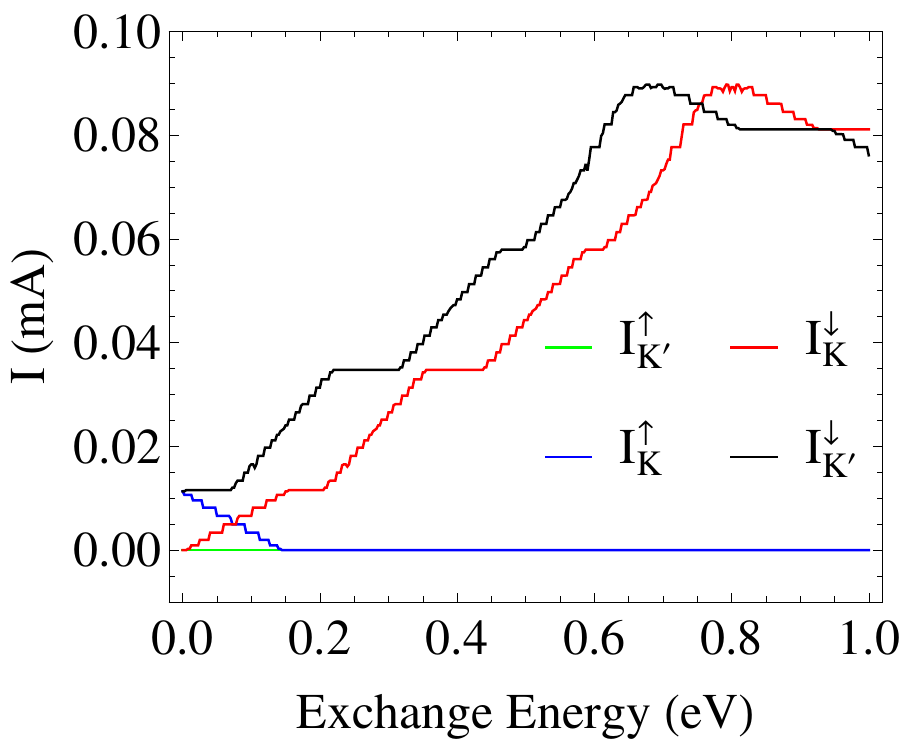}
\caption{(Color online) Current versus the exchange field. Here, $-0.9\le E\le -0.75$ eV, $h_{1}=h_{2}=M$ and $T=300$ K.}
\label{fig:Fig5}
\end{figure}

By sandwiching the channel between two magnetic insulators, the exchange energies induced by the top and the bottom magnetic insulators into A- and B-sublattices. It means that an exchange field, $h_{1}=h_{2}=M$, is applied to the carriers when they move between the leads. The configuration is called parallel configuration. Fig.~\ref{fig:Fig5} shows the plot of the current versus the exchange field for $-0.9\le E\le -0.75$ eV. As the figure shows, the valley polarization is equal to zero for $M=0$ and is equal to 46\% for $M=0.13$ eV. The valley polarization is only produced by spin-down carriers belonging to both $K$ and $K'$ valleys (see Fig.~\ref{fig:Fig5}).  

\begin{figure}	
\includegraphics[width=.8\linewidth]{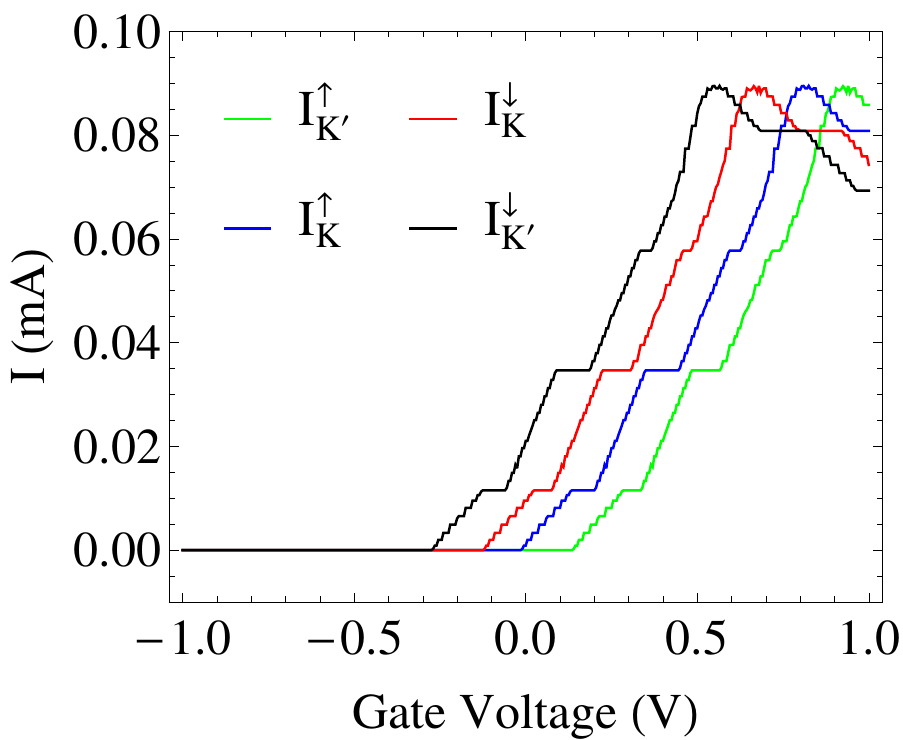}
\caption{(Color online) Current versus the gate voltage. Here, $-0.9\le E\le -0.75$, Here, $h_{1}=h_{2}=M=0.13$ and $T=300$ K.}
\label{fig:Fig6}
\end{figure}

However, the use of potential barriers is one of the common ways to generate a filter function. We should filter one type of valley DOF, for finding a perfect valley polarization. We can apply a gate voltage to the device, and adjust the difference between Fermi energy of leads and channel i.e., to adjust the height of potential barrier. If we apply a gate voltage to device and calculate the valley polarization for $-0.9\le E\le -0.75$ eV and $h_{1}=h_{2}=M=0.13$ eV, we will see a perfect polarization in the voltage range $ - 0.25 < {V_G} < - 0.11$ V as Fig.~\ref{fig:Fig6} shows. Here, this effect is created by the spin-down electrons belonging only to the $K'$ point.

Of course, it is highly desirable to obtain controllable and dual spin or valley polarization. Fig.7 (a)-(c) show the current versus gate voltage for three situations $h_{1}=M=-h_{2}$, $h_{2}=M=-h_{1}$, and $h_{1}=h_{1}=-M$, respectively ($M=0.13$). The situation $h_{1}=-h_{2}$ is called anti-parallel configuration. As Fig. 7(a) shows, the perfect valley polarization is created by the spin-up electrons belonging only to the $K$ point. Therefore, by switching the configuration from parallel to anti-parallel and measuring the type of spin, one can recognize the valley degree of freedom. Of course, from valley polarization point of view, Fig. 7(b) and Fig. 7(c) is similar to the Fig.6 and Fig. 7(a), respectively. Therefore, a particular spin-valley combination can be selected based on a given set of exchange field and gate voltage as input parameters. 

\begin{figure}[b]
\includegraphics{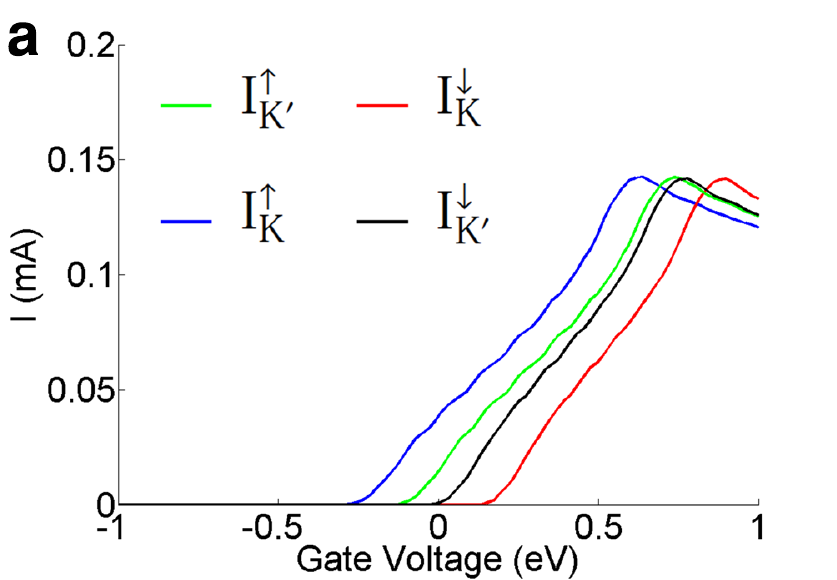}
\includegraphics{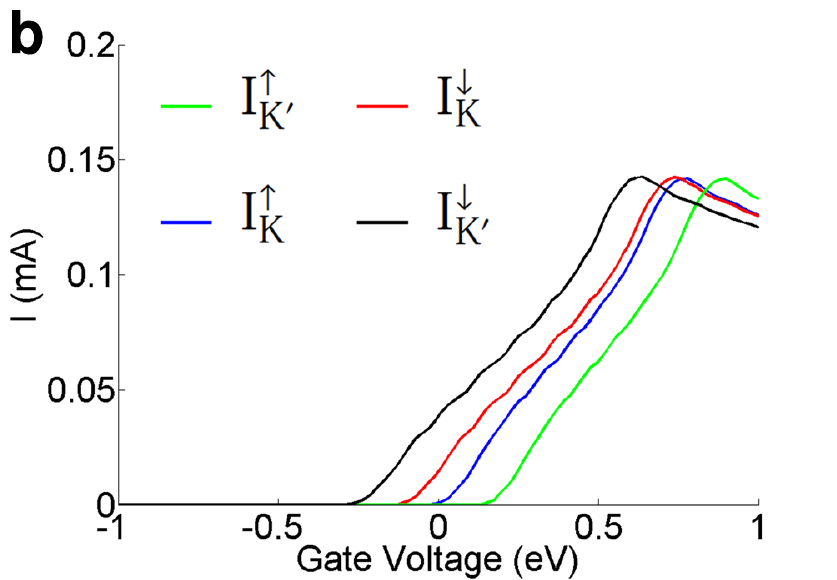}
\includegraphics{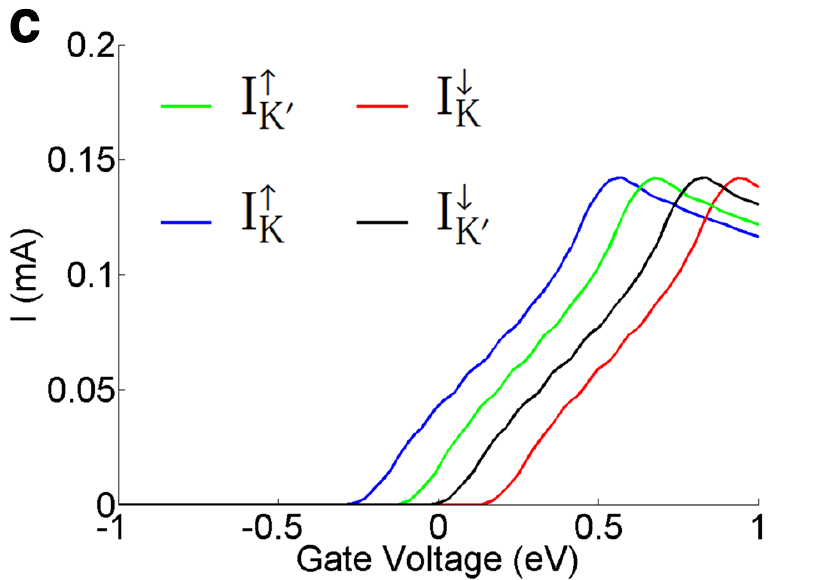}
\caption{(Color online) Current versus the gate voltage. Here, $-0.9\le E\le -0.75$, (a)$h_{1}=M=-h_{2}$, (b) $h_{2}=M=-h_{1}$, and (c) $h_{1}=h_{2}=-M$. Also, $M=0.13$ and $T=300$ K.}
\label{fig:Fig7}
\end{figure}

It is well known that, the number of transport channels increase when the width of nanoribbon increases, and in consequence, the conductance of nanoribbob increases. Therefore, it is expected that, the valley polarization be robust against the increasing the width of nanoribbon and its effects only appears in the level of current values (see Fig. 8).  

\begin{figure}
\includegraphics{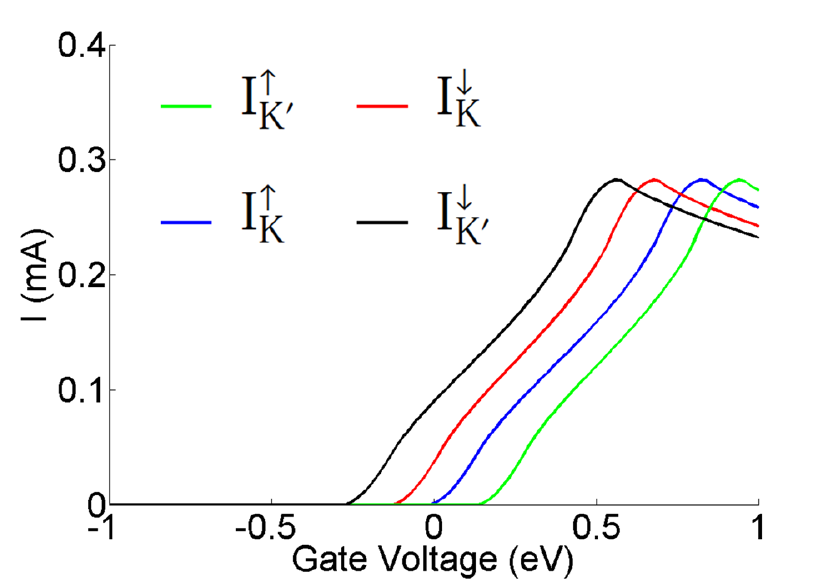}
\caption{(Color online) Current versus the gate voltage. Here, $-0.9\le E\le -0.75$, $h_{1}=h_{2}=M=0.13$ and $T=300$ K. Each unit cell includes 16 sites A and 16 sites B.}
\label{fig:Fig8}
\end{figure}

It should be noted that our structure is similar to the NM/SFM/NM structure introduced by Soodchomshom et al.\cite{ Soodchomshom 2014}.They have shown that the perfect spin-valley polarization only occurs when an anti-parallel (AP) junctions (i.e., $h_{1}=-h_{2}$) are considered and the chemical potential $(V_{G})$ is not equal to zero. Also, in their model, there is only one tunable variable i.e., an external electric field which is perpendicular to the plane of silicene. We showed perfect valley polarization can occur when $h_{1}=h_{2}$ and/or $h_{1}=-h_{2}$ and $V_{G}$ are not equal to zero. The tunable variables in our model are exchanged field and gate voltage. Yesilyurt et al. \cite{ Yesilyurt 2016} have used both strain and magnetic fields for removing the valley degeneracy and selecting the desired valley, respectively. They have found the perfect valley polarization with high conductance by changing the tunable variables, which were strain and magnetic gauge potentials.  Their strain and magnetic gauge fields look like our gate voltage and exchanged field, respectively. Of course, we used NEGF method while they (both Soodchomshom et al., and Yesilyurt et al.,) have used {\color{red} Dirac theory and done the analytical calculations.}

Finally, the robustness of perfect valley polarization against the disorder effects from nonmagnetic and magnetic impurities is important. For justifying the robustness, at least we should have the hopping integral between impurities and A- and B-sites of host atoms. Since we have not the hopping integral, we are not able to study the effect, here. However, it is an interesting subject, which can be studied in future by using a mixed method based on density functional theory and $\vec{k}.\vec{p}$ Hamiltonian model.

\section{Summary}\label{sec:summary}
We have studied the valley transport and valley polarization in zigzag nanoribbon of MoS$_2$ monolayer using the ${\vec k}.{\vec p}$ Hamiltonian model and the non-equilibrium Green's function method. It has been shown that, the quantum conductance plots versus electron energy include a transmission gap equal to 1.66 eV, perfect valley polarization for each spin degree of freedom at the edge of the valence band and no valley polarization when both spins are taken into account, due to the time-reversal symmetry of the Hamiltonian. We have shown that, an insignificant valley polarization was created by applying a magnetic field ($\vec B = {B_0}\hat z$ in which $B_0=30$ Tesla) to Mo-plane due to the breaking of the time-reversal symmetry. Also, by applying an exchanged field in parallel configuration ($\vec M = {M_0}\hat z$ where $M_0=0.13$ eV) a valley polarization equal to 46\% has been found which is only created by spin-down electrons belonging to both valleys. Finally, perfect valley polarization was created by applying a gate voltage when $M_0=0.13$ eV for both parallel and anti-parallel configurations. In parallel configuration spin-down is locked to $K'-point$ and in anti-parallel configuration spin-up is locked to $K-point$. Therefore, the valley polarization could be detected by detecting the spin degree of freedom.

\end{document}